\documentstyle[12pt,epsfig]{article}

\def\RMP{Rev. Mod. Phys.}
\def\ZP{Z. Phys.}
\def\PR{Phys. Rev.}
\def\RPP{Rep. Prog. Phys.}
\def\ANP{Ann. Phys.}
\def\EPC{{Eur. Phys. J.} C}
\def\NIMA{Nucl. Inst. and Meth. A}
\def\NPS{Nucl. Phys. B (Proc. Suppl.)}

\def\am{{$a_{\mu}$} }
\def\g2{{$(g-2)$}}
\def\ms{{$\mu$s}}
\def\bea{\begin{eqnarray}}
\def\eea{\end{eqnarray}}
\def\bc{\begin{center} }
\def\ec{\end{center} }

\def\beq{\begin{equation}}
\def\eeq#1{\label{#1}\end{equation}}
\def\eeqn{\end{equation}}

\def\beqa{\begin{eqnarray}}
\def\eeqa#1{\label{#1}\end{eqnarray}}
\def\eeqan{\end{eqnarray}}

\def\Journal#1#2#3#4{{#1} {\bf #2}, #3 (#4)}

\def\NPB{ Nucl. Phys. B}
\def\PLB{Phys. Lett.  B}
\def\PRL{Phys. Rev. Lett.}
\def\PRD{Phys. Rev. D}
\def\PR{Phys. Rev.}

\let\bar=\overbar

\def\Dslash{\not{\hbox{\kern-4pt $D$}}}
\def\dslash{\not{\hbox{\kern-2pt $\del$}}}

\def\msb{{\bar{\ssstyle M \kern -1pt S}}}

\def\Title#1{\begin{center} {\Large {\bf #1} } \end{center}}

\begin{document}

\Title{Status of the Muon \g2 Experiment  }

\bigskip\bigskip


\begin{raggedright}  

{\it B. Lee Roberts\index{Author, A.B.}\\
for the Muon \g2 Collaboration\cite{colab} \\
Department of Physics \\
Boston University,
Boston, Massachusetts 02215 }
\bigskip\bigskip
\end{raggedright}

\section{Introduction}

%

The study of $g$-factors of subatomic particles
can trace its roots back to the 1921
paper by Stern\cite{stern} which was a proposal to study space quantization
with an apparatus which is now called a ``Stern-Gerlach apparatus''. 
By 1924 the famous experiments had been done\cite{sg1}
and  a review paper was written summarizing their results.\cite{sg2}
Their final conclusion, that ``to within 10\%
the magnetic moment of the electron was one Bohr magneton'', meant
in modern language that the $g$-value of the electron was 2, where
the gyromagnetic ratio $g$ is the proportionality constant
between the magnetic moment and the
spin,
\beq
\vec \mu = g ({ e \over 2m}) \vec s.
\eeqn  

\begin{figure}[htb]
\begin{center}
\epsfig{file=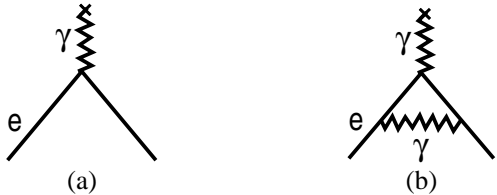,height=2.5cm}
\caption{The Feynman graphs for $g=2$ and the lowest
order radiative correction (Schwinger term)}
\label{fig:geq2sch}
\end{center}
\end{figure}

The discovery 
that $g_e \neq 2$,\cite{kusch}  and the calculation
by Schwinger\cite{schwinger} 
predicting that (to first order) the radiative correction to
 $g_e$ was $\alpha/\pi$, were important early steps in the development
of Quantum Electrodynamics (see Fig. \ref{fig:geq2sch}).

The long lifetime of the muon permits a precision measurement of 
its anomalous moment at the ppm level.
The muon magnetic moment is given by\cite{pdg}
\beq 
\mu_{\mu} = (1 + a_{\mu}) {e \hbar \over 2 m_{\mu}} \qquad {\rm where}
\qquad a_{\mu} = {(g-2)\over 2}.
\eeq{eq:mom}

The electron anomalous magnetic moment has been measured
to a few parts per
billion,\cite{vd} 
which can be completely described by QED of electrons and photons
to eighth order, $({\alpha\over \pi})^4$.
The contributions of virtual muons, tauons,
etc. enter at the few ppb level. The calculation of
the electron anomalous moment is limited by the knowledge of the
fine-structure constant.   With the reliability of modern QED
calculations, Kinoshita
has turned things around  and has used the electron $g$ value
measurement to give the
best value for $\alpha$.\cite{kinalpha}  

The relative contribution of heavier particles to the muon anomaly scales as
$( m_{\mu}/m_e )^2$ and the famous CERN experiment,\cite{cern3}
which obtained a relative error on \am of $\pm 7.3$ parts per million (ppm),
easily observed the 
predicted $\sim 60$ ppm contribution of
virtual hadrons.

In 1984 efforts began to make a new measurement of the muon anomalous
moment to a precision of $\pm 0.35$ ppm, which would represent a
5 standard deviation observation of the electroweak contribution, 
and would also be sensitive to contributions from ``new physics'' such
as muon substructure or supersymmetry.

\section{Theoretical Contributions to \g2}

The standard model value of \am is given by
$a_{\mu}{\rm ( SM)} 
= a_{\mu}({\rm QED}) + a_{\mu}({\rm hadronic}) + 
a_{\mu}({\rm weak})$
and any contribution from new physics would be reflected in a 
measured value
which did not agree with the standard model.

Comparison of the
measurements and calculations of the electron $g$ value gives one great 
confidence in our understanding of QED\cite{kinalpha} to the level
needed for muon \g2.
Taking the value of $\alpha$ from the electron 
$(g-2)$,\cite{kinalpha}
yields the total QED contribution
$a_{\mu}({\rm QED})=116\ 584\ 705.7(1.8)(0.5)\times 10^{-11}$.\cite{kinhughes}

\begin{figure}[htb]
\begin{center}
\epsfig{file=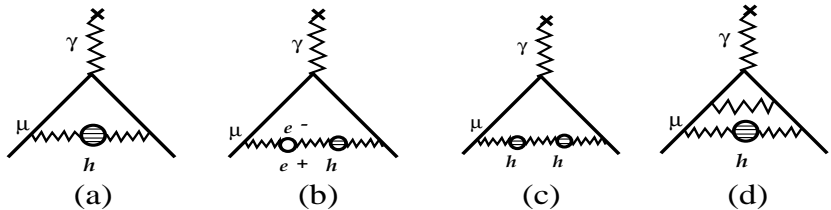 ,height=2.7cm,width=11cm}
\caption{(a) The lowest-order hadronic contribution. (b-d) Higher
order hadronic contributions except for the light-by-light
scattering contribution.
}
\label{fig:had}
\end{center}
\end{figure}

\begin{figure}[htb]
\begin{center}
\epsfig{file=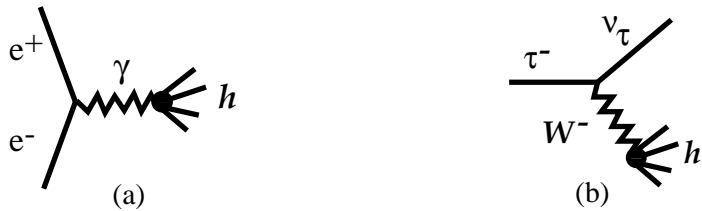,height=2.7cm}
\caption{(a). The hadroproduction process which enters the dispersion 
relation. (b) Hadronic $\tau$ decay.
}
\label{fig:hadpro}
\end{center}
\end{figure}

\begin{figure}[htb]
\begin{center}
\epsfig{file=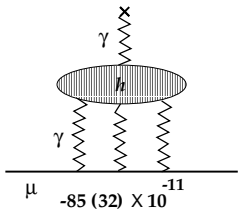,height=2.7cm}
\caption{(a). The hadronic light-by-light scattering contribution.
}
\label{fig:hadlol}
\end{center}
\end{figure}

The hadronic contribution to \g2 cannot be calculated directly, but
must be determined from data.  The first-order 
hadronic vacuum polarization dominates the uncertainty in the theoretical
value of $a_{\mu}$, since it is 
calculated using dispersion theory and
data from $e^+ e^- \rightarrow\  {\rm hadrons}$ and hadronic $\tau$ decay
as input.  The various order hadronic contributions are shown in 
Fig. \ref{fig:had}.  Diagrams for hadroproduction and hadronic $\tau$ 
decay are shown in Fig. \ref{fig:hadpro}.
The most precise determination of the first-order hadronic contribution
is\cite{dh}
$ a_{\mu}({\rm had};1)  = 6924 (62) \times 10^{-11}$ which is
 $59.39 \pm 0.53$ ppm of $a_{\mu}$, but there is continuing discussion of
the use of CVC and the $\tau$-decay data.\cite{ei}
The higher-order contribution is\cite{krause}
$a_{\mu}({\rm had};2)  = -101 (6) \times 10^{-11}$
The hadronic light-by-light scattering shown in Fig. \ref{fig:hadlol}
has now been calculated by two 
groups,\cite{hk,bpp} using essentially the same model, and
agreement is found: 
$ a_{\mu}({\rm had;lbl})  = -85 (32)10^{-11}$.
However the two groups
disagree on the uncertainty on the calculation, and I
have taken the larger error from Ref. \cite{bpp}.  The uncertainty
in this contribution could be reduced substantially by the appropriate
calculation on the lattice, and perhaps by other additional calculations
as well.\cite{deraph}

The total hadronic contribution is given by
$a_{\mu}({\rm had};1+2+{\rm lbl})  = 6738 (70) \times 10^{-11} $
which is $57.79 \pm 0.60$  ppm of $a_{\mu}$, with an uncertainty
dominated by the uncertainty on the first-order hadronic 
vacuum polarization.  It is precisely this contribution which
is being addressed by the programs to measure $R(s)$
at BES and the Budker
Institute.\cite{zz}  We look forward to additional high quality data
from these experiments, from DAPHNE,\cite{daphne} as well as $\tau$-decay
data from CLEO to further reduce the uncertainty on the hadronic 
contribution.

\begin{figure}[htb]
\begin{center}
\epsfig{file=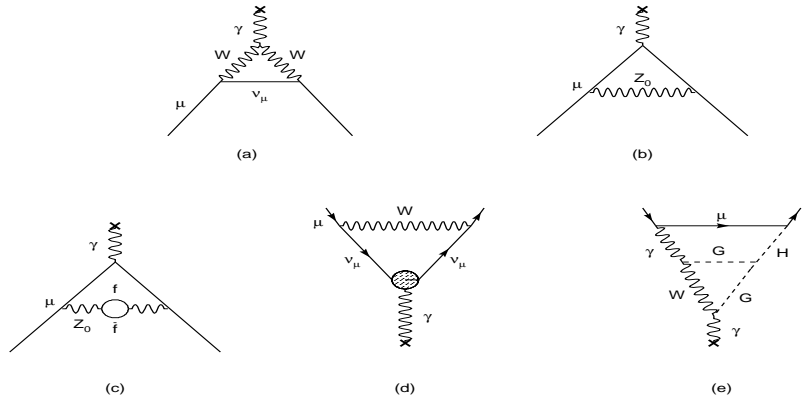,height=5.5cm,width=11.5cm}
\caption{(a,b). The single-loop electroweak contributions. (The standard
Higgs contribution is negligible.)
(c-e) Examples of the higher order electroweak contributions. 
(See Refs. \cite{BK1,BK2}.)
}
\label{fig:ew}
\end{center}
\end{figure}

The standard model electroweak contribution arises from the diagrams
shown in Fig. \ref{fig:ew} (the standard model Higgs contribution
is negligible).  
The single-loop $W$ and $Z$ contributions 
were calculated by a number of authors
shortly after the standard model was 
developed.\cite{BGL, JW, BY, deraf2}
The result is
$a_{\mu}({\rm weak };1) = 195 \times 10^{-11}$
or 1.7  ppm of $a_{\mu}$.
Partial calculations\cite{2W, 2W2} of the two-loop electroweak
contributions indicated that
they might not be small.  The full calculation\cite{BK1, BK2} which
was later confirmed independently\cite{Degr}
showed that the total first and second-order
weak contribution was 20\% less than  the first order result.  The result is
$a_{\mu}({\rm weak};1+2) = 151 (4) \times 10^{-11}$ which is
$1.30 \pm 0.03$  ppm of $a_{\mu}$.

The standard model prediction for \am is 
$a_{\mu} ({\rm SM}) = (116\ 591\ 594.7 \pm 70) \times 10^{-11}$
 ($\pm 0.60$ ppm).

A great deal has been written about the possible contribution to 
the muon \g2 value from non-standard model physics. Just as the proton
substructure produces a $g$-value which is not equal to two, muon
substructure would also contribute to the anomalous moment, the critical
issue being the scale of the substructure.\cite{sub}  A standard model
value for \g2 at the 0.35 ppm level would restrict the substructure scale
to around 5 TeV.

In Fig. \ref{fig:ew}(a) the triple gauge vertex $WW\gamma$ appears, and
it is through this diagram that the muon \g2 obtains its sensitivity to
$W$ substructure and anomalous gauge couplings.  The combined
sensitivity of LEP1, LEP2 and \g2, and the unique contribution 
which \g2 makes in constraining the existence of
such couplings, is described 
by Renard et al.\cite{anomcoup}

\begin{figure}[htb]
\begin{center}
\epsfig{file=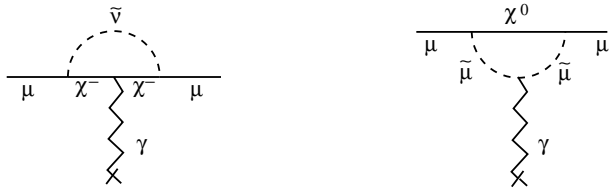,height=2.5cm}
\caption{
The lowest order supersymmetric contributions to \g2.
}
\label{fig:susy}
\end{center}
\end{figure}

Supersymmetry has become a serious candidate for physics beyond the standard
model.  The SUSY contribution is shown in 
Fig. \ref{fig:susy}.
In the case of large $\tan \beta$, the chargino diagram dominates
and the contribution to \g2 from SUSY is given by\cite{cm}
\beq
 a_{\mu}({\rm SUSY}) 
\simeq 
{\alpha \over 8 \pi \sin^2 \theta_W} {m_{\mu}^2 \over \tilde m^2} \tan \beta 
\simeq { 140 \times 10^{-11} } \  ({100\ {\rm GeV} \over \tilde m})^2
\tan \beta .
\eeq{eq:susy}
The goal of E821 is to reach a precision of
$\pm 40\times 10^{-11}$ ($\pm 0.35$ ppm), so the factor of
140 above corresponds to 1.2  ppm.  For $\tilde m= 750$ GeV 
and $\tan \beta = 40$,
$a_{\mu}({\rm SUSY}) = 100\times 10^{-11} $.

\section{The New \g2 Experiment}
For polarized muons moving in a uniform 
magnetic field $\vec B$ which is
perpendicular to the muon spin direction and to
the plane of the orbit, and
 with an electric quadrupole field $\vec E$ for
vertical focusing,\cite{cern3}  the difference angular frequency, $\omega_a$,
between the spin precession frequency $\omega_s$ and the cyclotron frequency
$\omega_c$, is given by
\beq
\vec \omega_a = - {e \over m }\left[ a_{\mu} \vec B -
\left( a_{\mu}- {1 \over \gamma^2 - 1}\right)
\vec \beta  \times \vec E \right].
\eeq{eq:omega}
The dependence of $\omega_a$ on the electric field 
is eliminated by storing
muons with the ``magic'' $\gamma_\mu$=29.3, which 
corresponds to a muon momentum
$p_{\mu}$ = 3.09 GeV/$c$.  Hence measurement of $\omega_a$ and of $B$
determines $a_\mu$. 
At the magic gamma, the muon lifetime is 
$\gamma \tau =64.4$ \ms,
the \g2 precession period is 4.37 \ms, and for 
the central orbit radius of 7.11 m the cyclotron period is 149 ns.

The storage ring magnet is a superferric 700 ton, 14 m diameter
circular ``C''-magnet, with
the opening facing inward towards the ring center.  The field is excited
by three 14 m diameter superconducting coils which carry
$5.2$ kA from a low voltage power
supply to produce the $1.45$~T
magnetic field.\cite{danby} 
The short term field stability over several AGS cycles is better than
0.1 ppm.

The magnetic field which enters in Eq. \ref{eq:omega} is the average field
seen by the muon distribution.
Since direct injection of muons does not uniformly fill the phase space,
we used a tracking code to calculate the distribution of muons in the
storage ring.  The radial distribution obtained from this tracking code was
compared with the distribution obtained from observing the beam debunching in
the ring at early times.  The two distributions agreed quite well.
In the 1999 run, a straw-tube array was operational at one detector location
which provided information on the decay positron trajectories coming out of
the storage region. These data will permit us to reconstruct directly the 
muon spatial distribution in one section of the ring.

In 1998, direct muon injection into the storage ring was employed
for the first time.  
The AGS performance, the beamline and the inflector magnet were as described
in Carey et al.,\cite{carey} and  except for the muon
injection many of the experimental details
are the same as described there.
 The positive muon beam with the magic momentum
is formed by collecting the highest energy muons from pion decay in 
a 72 m long decay section of our beamline, which results in a muon 
polarization of 96\%.  The flux incident on the inflector magnet 
 was $ 2 \times 10^6$ per fill of the ring.

The 10 mrad kick needed to put the muon beam onto a stable orbit was achieved
with a currents, since usual magnetic kicker techniques would spoil the
precision magnetic field.
Three pulse-forming networks powered three 
identical 1.7 m long kicker sections consisting of parallel
plates on either side of the beam.
Current flowed down one side crossed over and flowed
back up the other side.  The kicker plate geometry and composition was
chosen to minimize eddy currents, and the eddy current
effect on the total field seen by the muons was less than 0.1 ppm 20 \ms
\ after injection.
The current pulse, which was formed by an under-damped
LCR circuit, had a peak current of 4100 A and a pulse base width of 400 ns.
Since the cyclotron period of the muon beam from the AGS was 149 ns, the
beam was kicked several times before the kicker pulse died out.
With muon injection,
the number of detected positrons per hour was increased by an
order of magnitude over pion injection.  Thus the use of a muon beam 
an order of magnitude less intense than the pion beam resulted
in a substantial increase in stored muons per fill of the ring, with
the injection related background reduced by about a factor of 50.

Positrons from  the in-flight  decay 
$\mu^+\rightarrow e^+ \nu_e \bar \nu_{\mu}$
are detected with Pb-scintillating fiber calorimeters
placed symmetrically at 24
positions around the inside of the storage ring\cite{det}. 
  The decay positron time spectrum is\cite{cern3,fp}
\beq
N_0e^{-t/{\gamma\tau}}
\left[1+A(E)\cos\left(\omega_a t+\phi(E)\right)\right].
\label{eq:fivep} 
\eeqn
The normalization constant $N_0$ and the  parity violating 
asymmetry parameter $A(E)$ depend on the energy threshold placed
on the positrons. The fractional statistical error on 
$\omega_a$ is proportional to
$A^{-1}  N^{-1 /2}_{e}$, 
where $N_e$ is the number of decay positrons detected above some energy
threshold.  For an energy threshold 
of 1.8 GeV, we measure $A$ to be $0.34$  consistent with
its theoretical value,\cite{fp}
which we attribute to the good calorimeter energy resolution 
($\sigma/ E = 10\%$ at 1 GeV)
and a scalloped vacuum chamber which minimizes pre-showering before 
the positrons reach the calorimeters.

The  photomultiplier tubes of the calorimeter were gated off before injection 
and when gated on, they
recovered to 90\% pulse height in $\leq 400$ ns, and reached 
full operating gain in several $\mu$s.  
With the reduced
flash following injection 
it was possible to begin counting as soon as 5 \ms\   after injection 
in the region of the ring 180$^{\circ}$ around from the injection point.

The calorimeter pulses were continuously
sampled by custom 400 MHz waveform digitizers (WFDs), 
which provided both timing
and energy information for the positrons.  
Both the NMR and WFD clocks were phase-locked to the same LORAN-C frequency
signal.
The waveforms were zero-suppressed, and stored in memory
in the WFD until the end of the AGS cycle.  Between AGS acceleration cycles
the WFD 
data were written to tape for off-line analysis, as were the calorimeter
calibration data and the magnetic field data.

A laser/LED calibration system was
used to monitor calorimeter time and gain shifts during the
data-collection period. 
Early-to-late
timing shifts over the first 200 \ms \ were on-average less than 20 ps,
which is needed to keep systematic timing errors smaller than 0.1 ppm.

For the offline analysis,
the detector response (waveform shape) to positrons was determined from our
data for each calorimeter.
These shapes were then fit to all pulses in the data to determine
a time, an amplitude and
a width parameter for each pulse.
Time histograms were formed for each detector.
These independent data sets  were
analyzed separately and were in  agreement ($\chi^2/\nu = 17.2/20$).

A completely blind analysis was performed on the data.  Arbitrary offsets
were put on the muon frequency and the proton frequency from the NMR 
probes.  Each offset was known by one person, making it impossible
to determine the actual value of $a_{\mu}$. Only when the analysis of both the
magnetic field and $\omega_a$ were completed were the offsets removed and
the new value of \am determined.

After the offsets were removed,
it was necessary to make two corrections to the frequency obtained from the
fitting.  
For muons with the ``magic'' momentum, $\omega_a$ is not affected by
the electric field.  For the ensemble of muons in our storage
ring  there
is a small electric field correction  to $\omega_a$ since not all muons are
at the magic momentum.  There is also a pitch 
correction because of the vertical betatron
oscillations\cite{cern3,fp}.  The sum of these two corrections
for these data is ($ 0.9 \pm 0.2$) ppm.

The dominant systematic errors 
which were reported in our first measurement\cite{carey}
have been completely eliminated. The remaining 
systematic errors are under study.  With many of
them approximated by upper limits, one obtains a total systematic error of
$\leq 1$ ppm, 
with the systematic error assigned to the magnetic field of 0.5 ppm.
Since the study of systematic errors is a source of much continuing 
work, we have chosen not to present a detailed list at this time.  

\begin{figure}[htb]
\begin{center}
\epsfig{file=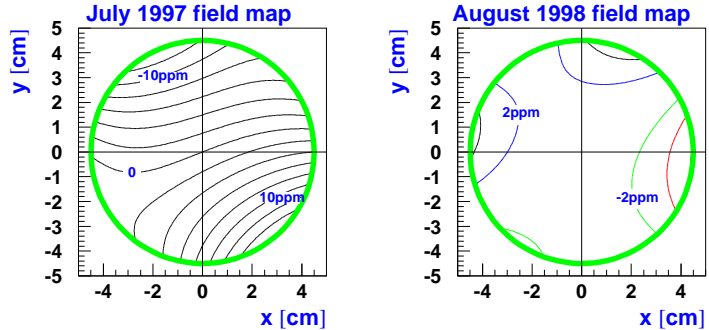,height=4.5cm}
\caption{ The magnetic field profile averaged over azimuth for July 1997
when the data reported in Carey et al.\cite{carey}
were taken,  and for August 1998
when the new data reported here were taken.  Each contour represents a 
2 ppm change.
}
\label{fig:field}
\end{center}
\end{figure}

We do wish to note the substantial improvement in the magnetic field quality
which has been obtained by additional shimming.  In Fig. \ref{fig:field} 
we show the average magnetic field from the 1997 and 1998 runs.
The field uniformity over the 
storage aperture in 1998 was almost an order of magnitude 
 better than was obtained in the CERN experiment.\cite{cern3}

\section{New Results}

One month after the Symposium,
we finished our analysis of the 1998 data and obtained a
new result at the precision of $\pm 5$ ppm.\cite{g2web} 

Our experiment measures the frequency ratio $R = \omega_a/\omega_p$, 
where $\omega_p$ is the free proton NMR frequency in our magnetic field.
Including the pitch and electric field corrections,
we obtain
$R= 3.707\ 201 (19) \times 10^{-3}$,
where the 5 ppm error includes a 1 ppm systematic error estimate.
We obtain $a_{\mu^+}$ from
$a_{\mu^+} = { R / ( \lambda - R) }$ $= 116\ 591\ 91 (59) \times 10^{-10}$
in which $\lambda = \mu_{\mu}/\mu_p = 3.183\ 345\ 39(10)$.\cite{pdg,lambda}
This new result is in good agreement with the mean of the CERN
measurements
for $a_{\mu^+}$ and $a_{\mu^-}$,\cite{pdg} and our previous measurement
of   $a_{\mu^+}$,\cite{carey} which are tabulated below.

\begin{table}[htb]
\begin{center}
\begin{tabular}{||l|l||}   \hline
{\em Measurement } & {\em Value} $\times 10^{10}$   \\
\hline
CERN\cite{cern3} $\mu^+$ & 116 591 03 (120) (10 ppm) \\
CERN\cite{cern3} $\mu^-$ & 116 593 65 (120) (10 ppm) \\
E821\cite{carey} $\mu^+$ $\pi_{\rm inj}$ & 116 592 51 (150) (13 ppm) \\
E821$^*$ $\mu^+$ $\mu_{\rm inj}$   & 116 591 91 \ (59)\ (\ 5 ppm) \\
\hline
New World Average & 116 592 10\ (46) (\ 4 ppm) \\
\hline
\end{tabular}
\caption{
The values of \am from the four most precise experiments.
The new value for $\mu_{\mu}/\mu_p$ has been used to get \am from
the measured ratio $\omega_a/\omega_p$. For the average,
 $\chi^2/\nu=0.92$.  The goal of the experiment is an error of
$\pm 4.0$ in the units above.
$^*$Preliminary }
\label{tab:results} 
\end{center}
\end{table}

\begin{figure}[!htb]
\begin{center}
\epsfig{file=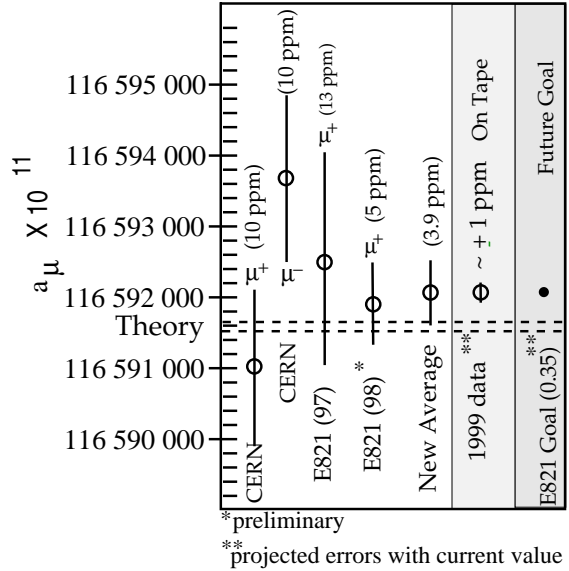,height=7.6cm}
\caption{The four precise measurements of \am and the standard model value.
To show the potential improvement available,
the projected errors of 1 ppm and 0.35 ppm are shown on the
right hand side with the central value equal to the present world average.
}
\label{fig:fiveppm}
\end{center}
\end{figure}

Assuming CPT symmetry,
the weighted mean of the four measurements
gives a new world average of
$a_{\mu} = 116\ 592\ 10 (46) \times 10^{-10}$ $(\pm 3.9\  {\rm ppm})$,
which agrees with the standard model to within one standard deviation.
These results are displayed graphically in Fig. \ref{fig:fiveppm}.
Also shown is the projected error from the 1999 data, and the $\pm 0.35$ ppm
goal of E821.

\section{Outlook and Conclusions}

The standard model has been remarkably successful in describing a wide range of
phenomena including the entire body of results from LEP.  The new result
from E821, which has  a precision of 5 parts per million, lowers the 
uncertainty on our knowledge of $(g-2)_{\mu}$ to 3.9 ppm.  
At that level there is
good agreement with the standard model, but there is still room for the
observation of a value at the level of one ppm or better which could 
agree with the previous measurements, and disagree with the standard model.
We expect our one ppm result to be available before the end of 2000.

This experiment is very much a work in progress.  The initial design of E821
was made with the systematic error budget of 0.12 ppm.  Many of the 
improvements which were made to the CERN technique have indeed worked, making
it straightforward to obtain a systematic error less than one ppm.  While
we have been learning about our systematic errors from the beginning of
the first pion injection run, with the sub-ppm statistics available with
muon injection we are now challenged to push the analysis of systematic
errors to the limit.
Our ultimate goal is a statistical answer at the 0.3 ppm level, with systematic
errors at perhaps half this level.
Our work thus far seems to indicate that 
this level of systematic error will be  possible. 

A new inflector magnet has been installed in the ring, which has
a fringe field in the storage region  a factor of 5 less
than the old inflector had.  This will improve our ability to map
the field everywhere and will help
to reduce the uncertainty
on our knowledge of the field.  The principal
magnetic field issue remaining is our ability
to track the field with time, and to understand the full calibration of the
NMR probes to a few tenths of a ppm.  

The other principal challenge is how to handle pile-up in the detectors without
discarding an unreasonable portion of the data set.  
The dominant analysis effort now is to understand this effect.

It has been 20 years since the CERN experiment presented its
final report  which verified the
hadronic contribution at the eight standard deviation level, but which
left it to a future experiment to verify the
electroweak contribution.  The new Brookhaven experiment is now approaching
that goal.  Within the next few years
either the standard model value will be confirmed, or evidence of a 
new contribution to the muon \g2 will be discovered.




\bigskip
I wish to acknowledge the efforts of the many collaborators who have worked
on \g2 over the past 15 years. The steady
support of the funding agencies (US and abroad) and the Brookhaven Laboratory
management was essential to our reaching this point.  I wish to thank 
R. Carey, D. Hertzog, K. Jungmann, V. Hughes, J. Miller, 
Y. Semertzidis and E. Sichtermann, for their
comments on this manuscript.


\begin{thebibliography}{99}
\bibitem{colab} The \g2 Collaboration consists of:
R.M. Carey, W. Earle, E. Efstathiadis, 
E.S. Hazen,  F. Krienen, I. Logashenko,
J.P. Miller, J. Paley, O. Rind, B.L. Roberts,
\ L.R. Sulak, A. Trofimov ( Boston University),
H.N. Brown, G. Bunce, G.T. Danby,
R. Larsen, Y.Y. Lee,
W. Meng, J. Mi, W.M. Morse, C. \"Ozben,  C. Pai, R. Prigl, R. Sanders,
Y. Semertzidis, D. Warburton (Brookhaven National Laboratory)
Y. Orlov (Cornell University)
D. Winn (Fairfield University),
A. Grossmann, K. Jungmann, G. zu Putlitz
(University of Heidelberg)
P.T. Debevec, W. Deninger, F. Gray
D.W. Hertzog, C. Onderwater, C. Polly, S. Sedykh, D. Urner
(University of Illinois)
U. Haeberlen (Max Planck Institiute f\"ur Med. 
Forschung, Heidelberg)
A. Yamamoto (KEK)
P. Cushman, L. Duong, S. Giron, J. Kindem, R. McNabb,
C. Timmermans, D. Zimmerman (University
of Minnesota)
V.P. Druzhinin, G.V. Fedotovich,
B.I. Khazin, N.M. Ryskulov, 
Yu.M. Shatunov, E. Solodov 
(Budker  Institute of Nuclear Physics, Novosibirsk),
M. Iwasaki, M. Kawamura (Tokyo Institute of Technology)
H. Deng, S.K. Dhawan, F.J.M. Farley, M. Grosse-Perdekamp,
V.W. Hughes, D. Kawall,  J. Pretz,
S.I. Redin, A. Steinmetz (Yale University) 

\bibitem{stern} O. Stern,  \Journal{\ZP}{7}{18}{1921}.

\bibitem{sg1}  W. Gerlach and O. Stern, \Journal{\ZP}{8}{110}{1922},
\Journal{\ZP}{9}{349}{1922} and \Journal{\ZP}{9}{353}{1924}.

\bibitem{sg2} W. Gerlach and O. Stern, \Journal{\ANP}{74}{45}{1924}.

\bibitem{kusch} J.E. Nafe, et al., \Journal{\PR}{71}{914}{1947},
D.E. Nagl, et al., \Journal{\PR}{72}{971}{1947}
P. Kusch and H.M. Foley, \Journal{\PR}{74}{250}{1948}.

\bibitem{schwinger} J. Schwinger, \Journal{\PR}{73}{416}{1948}
and \Journal{\PR}{75}{898}{1949}

\bibitem {pdg}
 Particle Data Group, 
\Journal{\EPC}{3}{1}{1998}

\bibitem{vd} R.S. Van Dyck et al., \Journal{PRL}{59}{26}{1987} and in
{\sl Quantum Electrodynamics}, 
(Directions in High Energy Physics Vol. 7) T. Kinoshita ed., World Scientific, 
1990, p.322.


\bibitem{kinalpha} T. Kinoshita, \Journal{\RPP}{59}{1459}{1996}.


\bibitem{cern3} J. Bailey, et. al,  \Journal{\NPB}{150}{1}{1979}.

\bibitem{kinhughes} V.W. Hughes and T. Kinoshita, 
\Journal{\RMP}{71}{S133}{1999}.

\bibitem{dh} M. Davier and A. H\"ocker,
\Journal{\PLB}{435}{427}{1998} and ref. therin.

\bibitem{ei} S.I. Eidelman and V.N. Ivanchenko, \Journal{\NPS}{76}{319}{1999}.

\bibitem{krause} Bernd Krause, 
\Journal{\PLB}{390}{392}{1997}.

\bibitem{hk} M. Hayakawa and T. Kinoshita, 
\Journal{\PRD}{57}{465}{1998} and ref. therein.

\bibitem{bpp} J. Bijnens, E. Pallante and J. Prades, 
\Journal{\NPB}{474}{379}{1996} and ref. therein.

\bibitem{deraph} E. de Raphael, private communication.

\bibitem{zz}Reviewed by  Zheng-guo Zhao, {\it R in Low Energy $e^+e^-$}, 
at this Conference.

\bibitem{daphne} Sergio Bertolucci, {\it Daone/Kloe Status Report}
at this Conference.

\bibitem{BGL} W.A. Bardeen, R. Gastmans and B Lautrup, 
\Journal{\NPB}{46}{319}{1972}.

\bibitem {JW} R. Jackiw and S. Weinberg, \Journal{\PRD}{5}{157}{1972}.

\bibitem {BY} I. Bars and M. Yoshimura,\Journal{\PRD}{6}{374}{1972}.
 
\bibitem{deraf2} J. Calmet, S. Narison, M. Perrottet and E. De Rafael,
\Journal{\RMP}{49}{21}{1977}.

\bibitem{2W} T.V. Kukhto, E.A. Kuraev, A. Schiller and Z.K.
Silagadze, \Journal{\NPB}{371}{567}{1992}.

\bibitem{2W2}S. Peris, M. Perrottet, E de Rafael, 
\Journal{\PLB}{355}{523}{1995}.

\bibitem{BK1}
A. Czarnecki, B. Krause and W.J. Marciano, 
\Journal{\PRD}{52}{R2619}{1995}.

\bibitem{BK2}A. Czarnecki, B. Krause and W.J. Marciano,
\Journal{\PRL}{76}{3267}{1996}.

\bibitem{Degr}G. Degrassi and G.F.
Giudice, \Journal{\PRD}{58}{53007}{1998}

\bibitem{sub}An overview of non-standard model physics is given by
 T. Kinoshita and W.J. Marciano in
{\em Quantum Electrodynamics} (Directions in High Energy Physics, Vol. 7),
ed. T. Kinoshita, (World Scientific, Singapore, 1990), p. 419.

\bibitem{anomcoup} F. Renard, et al., \Journal{\PLB}{409}{398}{1997}.

\bibitem{cm}A. Czarnecki and W. Marciano, 
\Journal{\NPS}{76}{245}{1999}.

\bibitem{danby} Gordon T. Danby, et. al., in preparation, and
G.T. Danby and J.W. Jackson,
Proceedings of the 1987 IEEE Particle Accelerator Conference, ed. 
E.R. Lindstron and L.S. Taylor, (IEEE, 1987) p. 1517.

\bibitem{carey} R.M. Carey et al., \Journal{\PRL}{82}{1632}{1999}.

\bibitem{meng} F. Krienen, D. Loomba and W. Meng, \Journal{\NIMA}{283}
{5}{1989}.

\bibitem {det} S. Sedykh et al, Proceedings of the VII Int. Conf. on
Calorimetry in High-Energy Physics (CALOR 97), Ed. E. Cheu, T. Embry, 
J. Rutherford and R. Wigmans, World Scientific (1998), p. 269.

\bibitem {fp} F.J.M. Farley and E. Picasso in {\em Quantum Electrodynamics},
ed. T. Kinoshita, (World Scientific, Singapore, 1990), p. 479.

\bibitem{g2web} see http://www.phy.bnl.gov/g2muon/home.html

\bibitem {lambda} W. Liu, et al., \Journal{\PRL}{82}{711}{1999}.


\end{thebibliography}
\end{document}